\documentclass[%
 showpacs,
 reprint,
 groupedaddress,
 showpacs,
 amsmath,amssymb,
 aps,
 prl,
]{revtex4-1}

\usepackage[pdftex]{graphicx, color}
\usepackage{bm}
\usepackage{braket}
\let\Re\relax
\let\Im\relax
\DeclareMathOperator{\Re}{Re}
\DeclareMathOperator{\Im}{Im}

\begin{document}

\title{Multipole Superconductivity in Nonsymmorphic Sr$_2$IrO$_4$}

\author{Shuntaro Sumita}
\email[]{s.sumita@scphys.kyoto-u.ac.jp}
\affiliation{%
 Department of Physics, Graduate School of Science, Kyoto University, Kyoto 606-8502, Japan
}%

\author{Takuya Nomoto}
\affiliation{%
 Department of Physics, Graduate School of Science, Kyoto University, Kyoto 606-8502, Japan
}%

\author{Youichi Yanase}
\affiliation{%
 Department of Physics, Graduate School of Science, Kyoto University, Kyoto 606-8502, Japan
}%


\date{\today}

\begin{abstract}
 Discoveries of marked similarities to high-$T_{\text{c}}$ cuprate superconductors point to the realization of superconductivity in the doped $J_{\text{eff}} = 1 / 2$ Mott insulator Sr$_2$IrO$_4$.
 Contrary to the mother compound of cuprate superconductors, several stacking patterns of in-plane canted antiferromagnetic moments have been reported, which are distinguished by the ferromagnetic components as $-++-$, $++++$, and $-+-+$.
 In this paper, we clarify unconventional features of the superconductivity coexisting with $-++-$ and $-+-+$ structures.
 Combining the group theoretical analysis and numerical calculations for an effective $J_{\text{eff}} = 1 / 2$ model, we show unusual superconducting gap structures in the $-++-$ state protected by nonsymmorphic magnetic space group symmetry.
 Furthermore, our calculation shows that the Fulde-Ferrell-Larkin-Ovchinnikov superconductivity is inevitably stabilized in the $-+-+$ state since the odd-parity magnetic $-+-+$ order makes the band structure asymmetric by cooperating with spin-orbit coupling.
 These unusual superconducting properties are signatures of magnetic multipole order in nonsymmorphic crystal.
\end{abstract}



\maketitle


A layered perovskite $5d$ transition metal oxide Sr$_2$IrO$_4$ has attracted recent attention because a lot of similarities to the high-temperature cuprate superconductors have been recognized.
For example, Sr$_2$IrO$_4$ (La$_2$CuO$_4$) has one hole per Ir (Cu) ion, and shows a pseudospin-$1 / 2$ antiferromagnetic order~\cite{BJKim2008}.
Moreover, recent experiments on electron-doped Sr$_2$IrO$_4$ indicate the emergence of a pseudogap~\cite{YKKim2014, Yan2015, Battisti2017} and at low temperatures a $d$-wave gap~\cite{YKKim2016}, which strengthens the analogy with cuprates.
Furthermore, $d$-wave superconductivity in Sr$_2$IrO$_4$ by carrier doping is theoretically predicted by several studies~\cite{Wang2011, HiroshiWatanabe2013, Yang2014, Meng2014}.
Distinct differences of Sr$_2$IrO$_4$ from cuprates are large spin-orbit coupling and nonsymmorphic crystal structure, both of which attract interest in the modern condensed matter physics.
In this Letter, we predict exotic superconducting properties in Sr$_2$IrO$_4$ unexpected in cuprates.

\begin{figure}[htbp]
 \centering
 \includegraphics[width=8cm, clip]{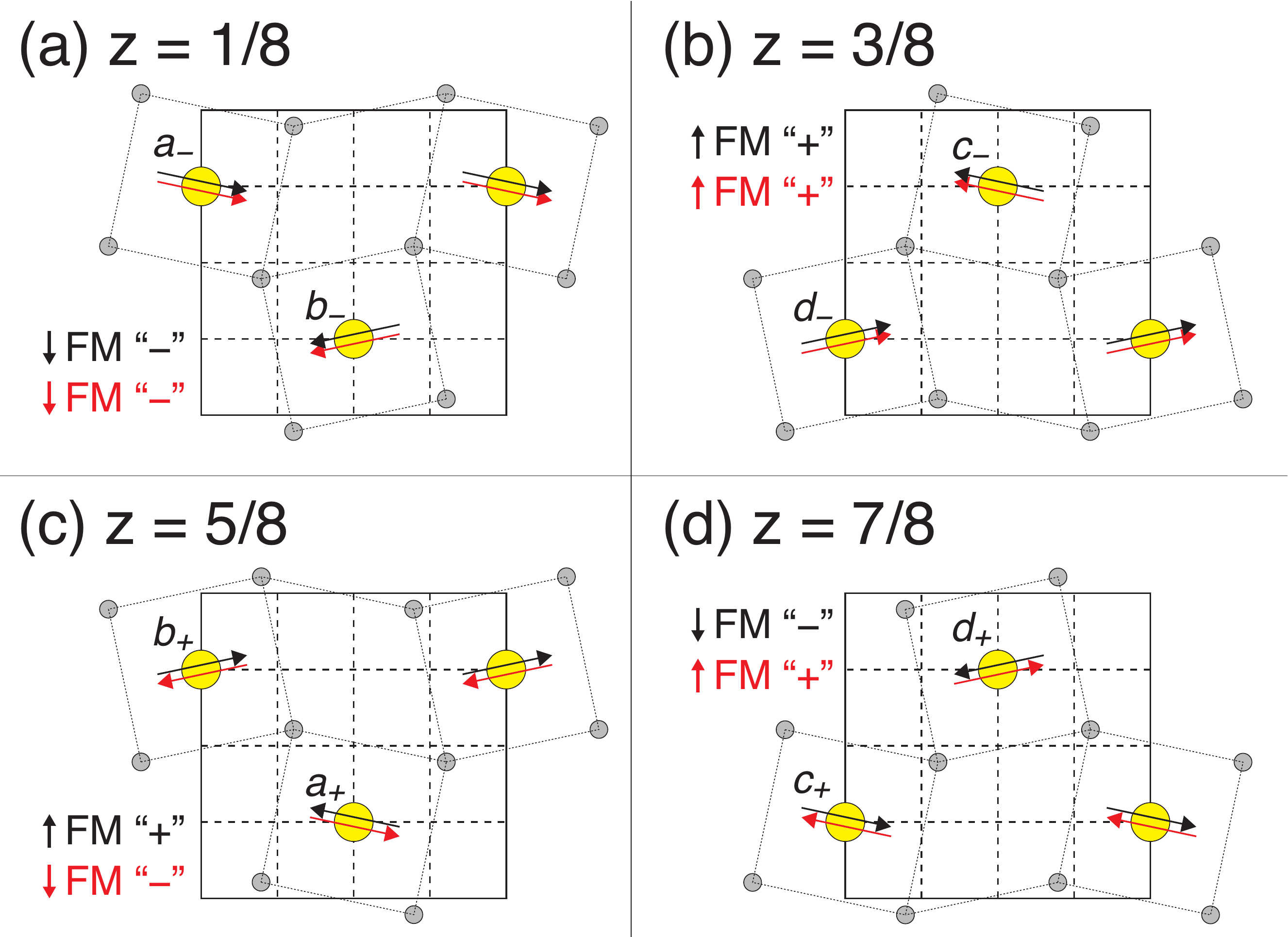}
 \caption{Crystal and magnetic symmetries of Sr$_2$IrO$_4$ in the 4 IrO$_2$ planes: (a) $z = \frac{1}{8}$, (b) $z = \frac{3}{8}$, (c) $z = \frac{5}{8}$, and (d) $z = \frac{7}{8}$~\cite{Matteo2016}. The two magnetic patterns of interest, $-++-$ (black arrows) and $-+-+$ (red arrows), are shown. They differ by the ferromagnetic in-plane component along the $b$ axis. Iridium atoms (yellow circles) are labeled as $a_-, \dots, d_-, a_+, \dots, d_+$.}
 \label{fig:Sr2IrO4}
\end{figure}

Below $T_{\text{N}} \simeq 230$ K, an antiferromagnetic order develops in undoped Sr$_2$IrO$_4$.
Large spin-orbit coupling and rotation of octahedra lead to canted magnetic moments from the $a$ axis and induce a small ferromagnetic moment along the $b$ axis (Fig.~\ref{fig:Sr2IrO4}).
Several magnetic structures for stacking along the $c$ axis have been reported in response to circumstances.
The magnetic ground states determined by resonant x-ray scattering~\cite{BJKim2009, Boseggia2013, Clancy2014}, neutron diffraction~\cite{Dhital2013, Ye2015}, and second-harmonic generation~\cite{Zhao2016}, are summarized in a recent theoretical work~\cite{Matteo2016}.
In the undoped compound, the ferromagnetic component shows the stacking pattern $-++-$~\cite{BJKim2009, Boseggia2013, Dhital2013}, as illustrated in Fig.~\ref{fig:Sr2IrO4}.
On the other hand, the $++++$ pattern is suggested as the magnetic structure of Sr$_2$IrO$_4$ in a magnetic field directed in the $ab$ plane~\cite{BJKim2009} and of Rh-doped Sr$_2$Ir$_{1 - x}$Rh$_x$O$_4$~\cite{Clancy2014, Ye2015}.
The recent observation~\cite{Zhao2016}, however, advocates the $-+-+$ magnetic pattern indicating an intriguing odd-parity hidden order in Sr$_2$IrO$_4$ (see Fig.~\ref{fig:Sr2IrO4}).

The crystal space group of Sr$_2$IrO$_4$ was originally reported as $I4_1/acd$ from neutron powder diffraction experiments~\cite{Huang1994, Crawford1994}.
Very recently, however, the crystal structure has been revealed by single-crystal neutron diffraction to be rather $I4_1/a$~\cite{Ye2015}.
In either case, the symmetry of Sr$_2$IrO$_4$ is globally centrosymmetric and nonsymmorphic.
On the other hand, the site symmetry of the Ir site is $S_4$ lacking local inversion symmetry.
In such noncentrosymmetric systems, antisymmetric spin-orbit coupling (ASOC) entangles various internal degrees of freedom, such as spin, orbital, and sublattice, namely multipole degrees of freedom.
As an intriguing consequence of the ASOC, locally noncentrosymmetric systems may realize odd-parity multipole order~\cite{Spaldin2008, Yanase2014, Hitomi2014, Hayami2014_02, Hayami2014_08, Hayami2015, LiangFu2015, Hitomi2016} beyond the paradigm of even-parity multipole order in $d$- and $f$-electron systems~\cite{Kuramoto_review}.

In noncentrosymmetric systems, exotic superconductivity such as the Fulde-Ferrell-Larkin-Ovchinnikov (FFLO) state~\cite{FF, LO} has been expected to be realized by the external magnetic field~\cite{Agterberg2007}.
Searches of the FFLO state have been an issue for more than five decades~\cite{Matsuda-Shimahara}.
For example, a recent experiment tries to detect a hallmark of the FFLO state in $\kappa$-(BEDT-TTF)$_2$Cu(NCS)$_2$~\cite{Mayaffre2014}.
However, it has been shown that in noncentrosymmetric systems the FFLO order parameter is hidden in vortex states~\cite{Matsunaga2008, Hiasa2009}.
Such difficulty of experimental researches may be resolved by odd-parity multipole order~\cite{Sumita2016}.
One of the purposes of this study is to propose material realization of the FFLO state free from disturbance by vortices.

Recent theories have shed light on mathematically rigorous properties ensured by nonsymmorphic crystal symmetry~\cite{Shiozaki2015, Fang2015, Po2015, HarukiWatanabe2015, HarukiWatanabe2016, Shiozaki2016}.
For nonsymmorphic superconductors, nodal-line superconductivity unexpected from existing classification based on the point group~\cite{Sigrist-Ueda} was found by Norman in 1995~\cite{Norman1995}.
Unconventional superconductivity possessing such symmetry-protected line nodes is expected to appear in UPt$_3$~\cite{Norman1995, Micklitz2009, Kobayashi2016, Yanase2016, Nomoto2016, Micklitz2017_PRB}, UCoGe~\cite{Nomoto2017}, and UPd$_2$Al$_3$~\cite{Fujimoto2006, Nomoto2017, Micklitz2017_PRL}, due to the effect of spin-orbit coupling or magnetic order.
However, nonsymmorphic superconductivity by multipole order has not been uncovered.

In this Letter, we show that Sr$_2$IrO$_4$ may be a platform realizing two unconventional superconducting states, assuming the coexistence with magnetic order~\cite{magnetic_order}.
First, superconductivity with nonsymmorphic symmetry-protected gap structures is induced by the $-++-$ order, which is regarded as a higher-order magnetic octupole (MO) order.
Second, the FFLO superconductivity free from vortices is stabilized in the $-+-+$ [magnetic quadrupole (MQ)] state.
These results are evidenced by a combination of group theoretical analysis and numerical analysis of an effective $J_{\text{eff}} = 1 / 2$ model for Sr$_2$IrO$_4$.

\textit{$-++-$ state ---}
Now we consider the superconductivity in the $-++-$ state.
We begin with the gap classification based on the space group (see the Supplemental Material~\cite{supplemental}).
The magnetic space group of the $-++-$ state, $M_{-++-}$, is a nonsymmorphic group $P_Icca$.
We especially focus on the Cooper pairs on the basal planes (BPs) $k_{z, x, y} = 0$ and the zone faces (ZFs) $k_z = \pm \pi / c$ and $k_{x, y} = \pm \pi / a$.
In these high-symmetry planes, the small representation $\gamma^{\bm{k}}_{-++-}$ can be calculated.
Indeed, $\gamma^{\bm{k}}_{-++-}$ corresponds to the Bloch state with the crystal momentum $\bm{k}$.

In the superconducting state, the zero-momentum Cooper pairs have to be formed between the degenerate states present at $\bm{k}$ and $- \bm{k}$ within the weak-coupling BCS theory.
Therefore, these two states should be connected by some symmetry operations, such as space inversion.
As a result, the representation of Cooper pair wave functions $P^{\bm{k}}_{-++-}$ can be constructed from the representations of the Bloch state $\gamma^{\bm{k}}_{-++-}$~\cite{Bradley, Mackey1953, Bradley1970}.

We here calculate the character of the representation $P^{\bm{k}}_{-++-}$, and then reduce $P^{\bm{k}}_{-++-}$ into irreducible representations (IRs) of the original crystal symmetry $D_{4h}$.
The obtained results are summarized in the following:
\begin{itemize}
 \item $k_z = 0, \, \pm \pi / c$
       \begin{equation}
        \begin{cases}
         \begin{aligned}
          A_{1g} &+ A_{2g} + B_{1g} + B_{2g} + 2 A_{1u} \\
          &+ 2 A_{2u} + 2 B_{1u} + 2 B_{2u} + 2 E_u
         \end{aligned}
         & \text{BP} \\
         2 E_g + A_{1u} + A_{2u} + B_{1u} + B_{2u} + 4 E_u & \text{ZF}
        \end{cases}
        \label{eq:gap_classification_horizontal}
       \end{equation}
 \item $k_{x, y} = 0, \, \pm \pi / a$
       \begin{equation}
        \begin{cases}
         \begin{aligned}
          A_{1g} &+ B_{1g} + E_g + 2 A_{1u} \\
          &+ A_{2u} + 2 B_{1u} + B_{2u} + 3 E_u
         \end{aligned}
         & \text{BP} \\
         A_{2g} + B_{2g} + E_g + 3 A_{1u} + 3 B_{1u} + 3 E_u & \text{ZF}
        \end{cases}
        \label{eq:gap_classification_vertical}
       \end{equation}
\end{itemize}
We find that possible IRs change from BPs to ZFs as a consequence of the nonsymmorphic symmetry.
The gap functions should be zero, and thus, the gap nodes appear, if the corresponding IRs do not exist in these results of reductions~\cite{Izyumov1989, Yarzhemsky1992, Yarzhemsky1998}.
Otherwise, the superconducting gap will open in general.
From Eqs.~\eqref{eq:gap_classification_horizontal} and \eqref{eq:gap_classification_vertical}, for instance, we find the gap structure of $A_{1g}$ and $B_{2g}$ superconducting states summarized in Table~\ref{tab:gap_structure}.

\begin{table}[htbp]
 \centering
 \caption{The gap structure for $A_{1g}$ and $B_{2g}$ gap functions.}
 \label{tab:gap_structure}
 \begin{tabular}{c|cccc} \hline\hline
  & $k_z = 0$ & $k_z = \pm \pi / c$ & $k_{x, y} = 0$ & $k_{x, y} = \pm \pi / a$ \\ \hline
  $A_{1g}$ ($s$-wave) & gap & node & gap & node \\
  $B_{2g}$ ($d_{xy}$-wave) & gap & node & node & gap \\ \hline\hline
 \end{tabular}
\end{table}

We demonstrate the results of group theory (Table~\ref{tab:gap_structure}) using a three-dimensional single-orbital tight-binding model for $J_{\text{eff}} = 1 / 2$~\cite{supplemental} manifold.
Eight Ir atoms per unit cell and three types of ASOC~\cite{ASOC} are taken into account.
We consider the $s$-wave order parameter~\cite{order_parameter} which belongs to the $A_{1g}$ representation of the point group $D_{4h}$,
\begin{equation}
 \hat{\Delta}^{(s)}(\bm{k}) = \Delta_0 \hat{1}_2 \otimes \hat{\sigma}_0^{\text{(layer)}} \otimes \hat{\sigma}_0^{\text{(sl)}} \otimes i \hat{\sigma}_y^{\text{(spin)}},
\end{equation}
and the $d_{xy}$-wave order parameter~\cite{order_parameter} which belongs to the $B_{2g}$ representation,
\begin{equation}
 \hat{\Delta}^{(d)}(\bm{k}) = \Delta_0 \sin\frac{k_x a}{2} \sin\frac{k_y a}{2} \hat{1}_2 \otimes \hat{\sigma}_0^{\text{(layer)}} \otimes \hat{\sigma}_x^{\text{(sl)}} \otimes i \hat{\sigma}_y^{\text{(spin)}},
\end{equation}
where $\hat{1}_M$ is a $M \times M$ identity matrix.
$\hat{\sigma}_i^{\text{(spin)}}$, $\hat{\sigma}_i^{\text{(sl)}}$, and $\hat{\sigma}_i^{\text{(layer)}}$ are the Pauli matrices representing the spin, sublattice, and layer degrees of freedom, respectively.

\begin{figure*}[htbp]
 \centering
 \includegraphics[width=.95\linewidth, clip]{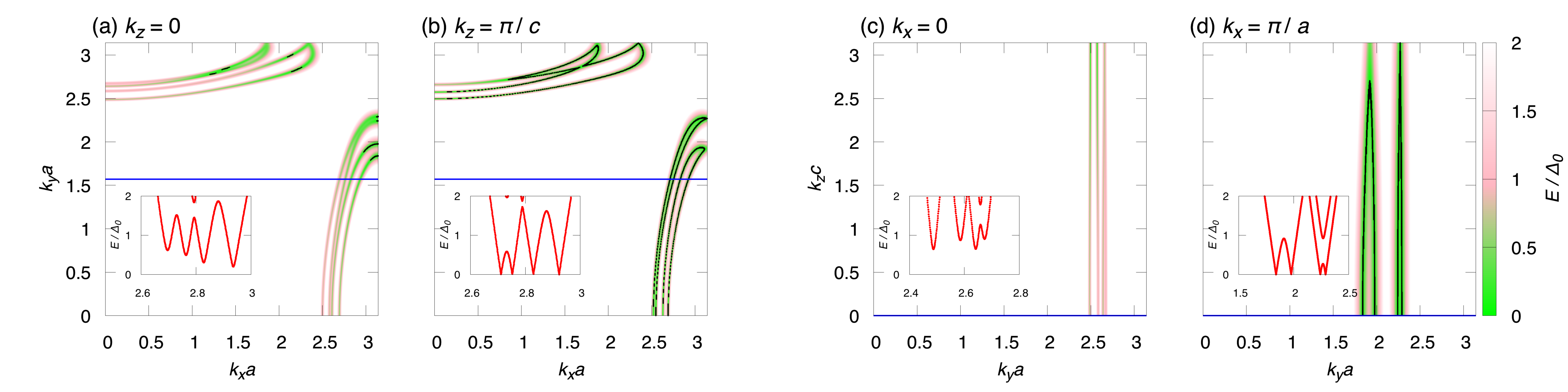}
 \caption{The contour plot of quasiparticle energy dispersion $E$ in the $s$-wave superconducting state normalized by the order parameter $\Delta_0$ on (a) $k_z = 0$, (b) $k_z = \pm \pi / c$, (c) $k_x = 0$, and (d) $k_x = \pm \pi / a$. The insets in (a), (b), (c), and (d) show the dispersion $E / \Delta_0$ along the respective blue line. Line nodes (black lines) appear on the ZF, $k_z = \pm \pi / c$ and $k_x = \pm \pi / a$.}
 \label{fig:mppm_s-wave}
\end{figure*}

\begin{figure*}[htbp]
 \centering
 \includegraphics[width=.95\linewidth, clip]{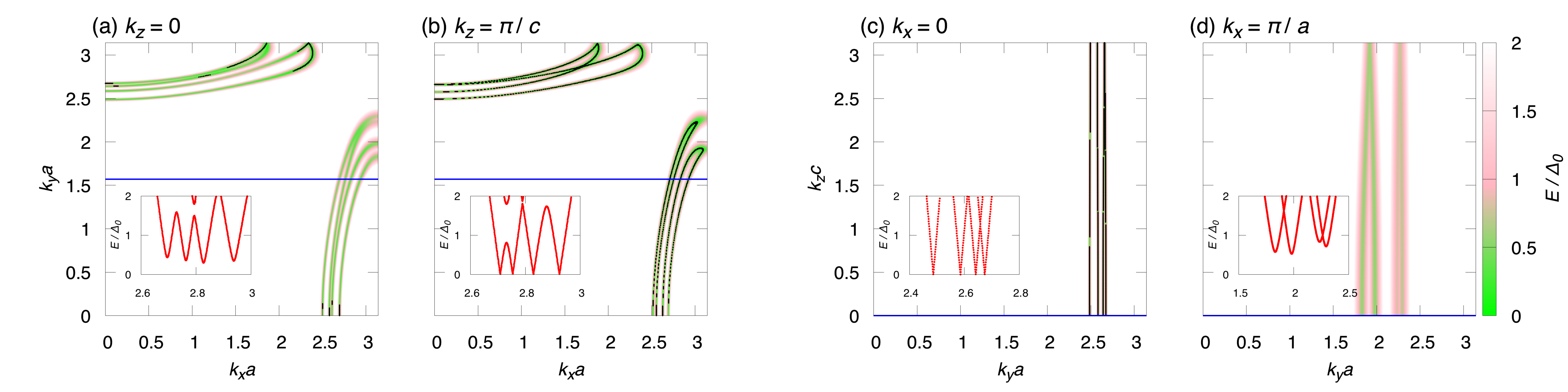}
 \caption{The contour plot of quasiparticle energy dispersion $E / \Delta_0$ for the $d_{xy}$-wave order parameter on (a) $k_z = 0$, (b) $k_z = \pm \pi / c$, (c) $k_x = 0$, and (d) $k_x = \pm \pi / a$. The insets show $E / \Delta_0$ along the respective blue line. Line nodes (black lines) appear on the ZF $k_z = \pm \pi / c$ and the BP $k_x = 0$.}
 \label{fig:mppm_d-wave}
\end{figure*}

The quasiparticle energy dispersion in the superconducting state $E = E(k_x, k_y, k_z)$ is obtained by diagonalizing the Bogoliubov-de Gennes (BdG) Hamiltonian~\cite{supplemental},
\begin{equation}
 \hat{H}_{\text{BdG}}(\bm{k}) =
  \begin{pmatrix}
   \hat{H}_{n}(\bm{k}) & \hat{\Delta}(\bm{k}) \\
   \hat{\Delta}(\bm{k})^\dagger & - \hat{H}_{n}^{\text{T}}(- \bm{k})
  \end{pmatrix}.
\end{equation}
The chemical potential is chosen to set the electron density $n \sim 1.2$, around which the superconductivity has been predicted~\cite{HiroshiWatanabe2013}.
However, superconducting properties revealed below are independent of the electron density.
The numerical results are shown in Figs.~\ref{fig:mppm_s-wave} and \ref{fig:mppm_d-wave}.
Only $0 \leq E / \Delta_0 < 2$ region is colored, and especially nodal ($E \sim 0$) points are plotted by black.

The gap structure of the two superconducting states reproduces Table~\ref{tab:gap_structure}.
In both $s$-wave and $d_{xy}$-wave cases, the numerical results are consistent with the group theory.
In other words, the gap nodes in Figs.~\ref{fig:mppm_s-wave} and \ref{fig:mppm_d-wave} are protected by nonsymmorphic space group symmetry.
Note that exceptional cases of the gap classification in Table~\ref{tab:gap_structure} appear in some accidentally degenerate region~\cite{Yanase2016}.
For example, we see such unexpected gap structures on the $k_y = \pm \pi / a$ plane~\cite{supplemental}.

As introduced previously, both theory~\cite{Wang2011, HiroshiWatanabe2013, Yang2014, Meng2014} and experiment~\cite{YKKim2016} suggest $d_{xy}$-wave superconductivity analogous to cuprates~\cite{d-wave}.
In this case, a horizontal line node appears on the ZF ($k_z = \pm \pi / c$) in contrast to the usual $d_{xy}$-wave state.
Moreover, the gap opening at the other ZFs ($k_{x, y} = \pm \pi / a$) is also nontrivial because the usual $d_{xy}$-wave order parameter vanishes not only at BPs but also at ZFs.
These nontrivial gap structures are protected by the nonsymmorphic space group symmetry.

\textit{$-+-+$ state ---}
We now turn to the $-+-+$ state of Sr$_2$IrO$_4$.
In this case, the method of gap classification used above is not applicable since there is no symmetry operation connecting $\bm{k}$ to $- \bm{k}$.
Conversely, Cooper pairs do not need to be formed between $\bm{k}$ and $- \bm{k}$ states, which indicates the emergence of the FFLO superconductivity.
Indeed, the FFLO state is stabilized in the $-+-+$ state as shown below.

Before going to the main result, here we show that the $-+-+$ order can be regarded as an odd-parity MQ order, which results in the asymmetry in the band structure.
Using a group theoretical analysis, it is determined that the $-+-+$ order belongs to $E_u$ representation of $D_{4h}$~\cite{supplemental}.
This IR permits time-reversal-odd basis functions: $\alpha y \hat{\sigma}_z + \beta z \hat{\sigma}_y$ in the real space, and $k_x$ in the momentum space.
In the real space, the basis function represents a rank-2 odd-parity MQ order~\cite{Schwartz1955},
\begin{equation}
 \hat{M}_{2, 1} + \hat{M}_{2, -1} \propto y \hat{z} + z \hat{y},
\end{equation}
where $\hat{M}_{l, m}$ is the magnetic multipole operator.
Therefore, the $-+-+$ order contains the component of a MQ order, though it may include a toroidal dipole order proportional to $y \hat{z} - z \hat{y}$~\cite{Spaldin2008}.
In the momentum space, the linear $k_x$ function makes the band structure asymmetric along the $k_x$ axis.
We actually confirm the asymmetry of the band structure using our tight-binding model~\cite{supplemental}.
Then, we also notice a twofold degeneracy in the band structure protected by symmetry~\cite{supplemental}.
These features of band structure resemble the MQ state in the zigzag chain~\cite{Yanase2014, Sumita2016}.
A similar analysis identifies the $-++-$ magnetic order as an even-parity MO order with $x y \hat{z} + y z \hat{x} + z x \hat{y}$.

Next, we study the superconductivity in the $-+-+$ state.
We can clarify the superconducting state near the transition temperature by linearizing the BdG equation while avoiding the numerical limitations of the full BdG equation.
The linearized BdG equation is formulated by calculating the superconducting susceptibility $\chi_{m m'}(\bm{q}, i \Omega_n)$~\cite{supplemental}, where $\Omega_n = 2 n \pi T$ is the bosonic Matsubara frequency, and $m$ represents the sublattice degrees of freedom.
Here we assume the local $s$-wave superconductivity for simplicity.
The $8 \times 8$ susceptibility matrix $\hat{\chi} = (\chi_{m m'})$ is obtained by the $T$-matrix approximation~\cite{TWatanabe2015},
\begin{equation}
 \hat{\chi}(\bm{q}, i \Omega_n) = \frac{\hat{\chi}^{(0)}(\bm{q}, i \Omega_n)}{\hat{1}_8 - U \hat{\chi}^{(0)}(\bm{q}, i \Omega_n)},
\end{equation}
where $U$ is the $s$-wave on-site attraction, and $\hat{\chi}^{(0)}$ is the irreducible susceptibility.

The superconducting transition occurs at the temperature $T_{\text{c}}$ where $\hat{\chi}(\bm{q}, i \Omega_n)$ diverges.
Thus, the criterion of the superconducting instability is $\chi^{(0)}_{\text{max}}(\bm{q}, i \Omega_n) = 1$, where $\chi^{(0)}_{\text{max}}$ is the largest eigenvalue of $U \hat{\chi}^{(0)}$.
Here $\chi_{\text{max}}^{(0)}$ shows the maximum at $q_y = q_z = \Omega_n = 0$, since energy bands are symmetric with respect to $k_y$ and $k_z$ even in the $-+-+$ state~\cite{supplemental}.

\begin{figure}[htbp]
 \centering
 \includegraphics[width=8cm, clip]{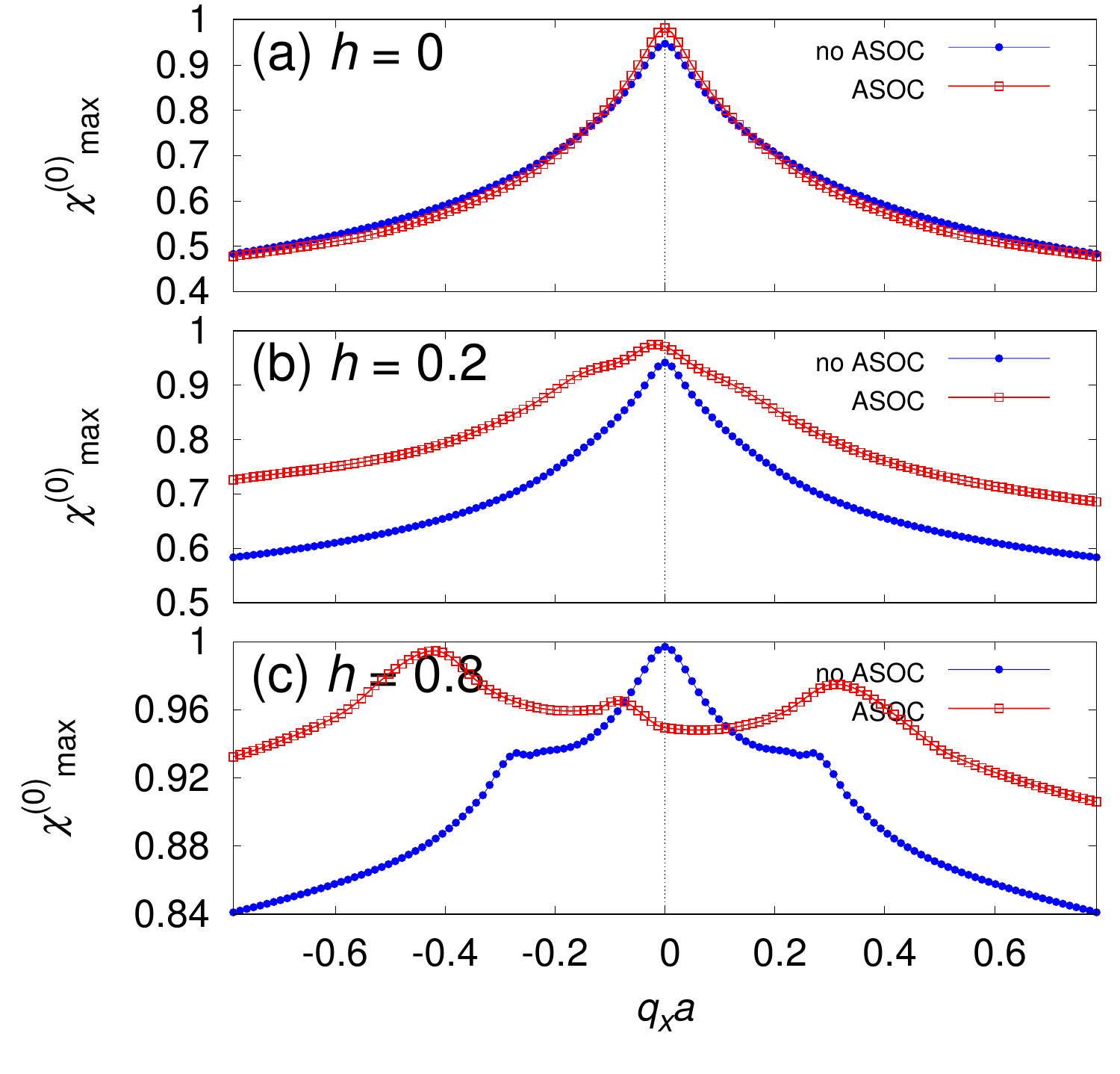}
 \caption{The largest eigenvalue $\chi^{(0)}_{\text{max}}$ in (a) the normal state ($h = 0$), (b) the small moment $-+-+$ state ($h = 0.2$), and (c) the large moment $-+-+$ state ($h = 0.8$). We fix the temperature $T = 0.01 \sim T_{\text{c}}$. For $h = 0$, $0.2$, and $0.8$, the $s$-wave on-site interaction $U$ is respectively assumed to be $0.26$, $0.47$, and $1.45$ in the absence of the ASOC, while it is $0.31$, $0.55$, and $1.60$ in the presence of the ASOC.}
 \label{fig:mpmp_sus}
\end{figure}

Figure~\ref{fig:mpmp_sus} shows the $q_x$ dependence of $\chi^{(0)}_{\text{max}}(\bm{q}, 0)$ at $T \sim T_{\text{c}}$.
In the normal state ($h = 0$), since the system preserves the inversion symmetry, $\chi^{(0)}_{\text{max}}$ has a peak at $q_x = 0$ regardless of the presence or absence of the ASOC [Fig.~\ref{fig:mpmp_sus}(a)].
On the other hand, in the $-+-+$ state ($h = 0.2$ and $0.8$), $\chi^{(0)}_{\text{max}}$ shows the maximum at a finite $q_x$ when the ASOC exists, while the conventional $\bm{q} = \bm{0}$ state is stable in the absence of the ASOC [Figs.~\ref{fig:mpmp_sus}(b) and \ref{fig:mpmp_sus}(c)].
This result reveals that the FFLO state is favored by the ASOC in the odd-parity $-+-+$ magnetic ordered state, despite the absence of the macroscopic magnetization required for the conventional FFLO state~\cite{FF, LO, Agterberg2007, Matsuda-Shimahara, Mayaffre2014}.
Moreover in the large moment state ($h = 0.8$), three local maxima are observed in Fig.~\ref{fig:mpmp_sus}(c).
The behavior resembles the band-dependent FFLO state in the one-dimensional zigzag chain~\cite{Sumita2016}.
Namely, a part of the bands mainly causes the superconductivity, while the other bands are weakly superconducting.
The nonuniform state with a large $|q_x a| \sim 0.4$ should be regarded as a pair-density-wave state~\cite{Agterberg2009, Wang2015, Freire2015} rather than the FFLO state.

\textit{Summary ---}
In this Letter, we investigated the superconductivity of doped Sr$_2$IrO$_4$ in the two magnetic states, $-++-$ and $-+-+$.
In the $-++-$ (MO) state, both $s$-wave and $d_{xy}$-wave superconductivity shows nontrivial line nodes protected by nonsymmorphic symmetry on the BZ boundary.
The nodal gap is analogous to that studied in toy models~\cite{Nomoto2016, Micklitz2017_PRB, Fujimoto2006}.
In a realistic model for Sr$_2$IrO$_4$, however, we have clarified not only nontrivial line nodes but also an unexpected gap opening.
In the case of $d_{xy}$-wave superconductivity, the gap opens on the vertical BZ face unlike the ordinary $d_{xy}$-wave superconductor.
On the other hand, in the $-+-+$ state identified as parity-violating odd-parity MQ state, the FFLO state is stabilized irrespective of the magnitude of the antiferromagnetic moment, because the band structure asymmetrically deforms.
The asymmetric band structure and resulting FFLO superconductivity are regarded as magnetoelectric effects caused by odd-parity MQ order.
The FFLO state caused by the MQ order does not need an external magnetic field, which means the ``pure FFLO state'', namely the FFLO state free from vortices.
Material realization in Sr$_2$IrO$_4$ may enable experimental observation of FFLO superconductivity.

We suggest doped Sr$_2$IrO$_4$ as a platform of nonsymmorphic nodal superconductivity by magnetic multipole order.
Furthermore, the realization of parity-violating multipole and FFLO superconductivity are proposed beyond the toy model~\cite{Sumita2016}.
These results point to nontrivial interplay of magnetic multipole order and superconductivity in the strongly spin-orbit coupled systems.

\begin{acknowledgments}
 The authors are grateful to H.~Watanabe, S.~Kobayashi, and M.~Sato for fruitful discussions.
 This work was supported by Grant-in Aid for Scientific Research on Innovative Areas ``J-Physics'' (15H05884) and ``Topological Materials Science'' (16H00991) from JSPS of Japan, and by JSPS KAKENHI Grants No. 15K05164, No. 15H05745, and No. 15J01476.
\end{acknowledgments}


%

\clearpage

\renewcommand{\thesection}{S\arabic{section}}
\renewcommand{\theequation}{S\arabic{equation}}
\setcounter{equation}{0}
\renewcommand{\thefigure}{S\arabic{figure}}
\setcounter{figure}{0}
\renewcommand{\thetable}{S\arabic{table}}
\setcounter{table}{0}
\makeatletter
\c@secnumdepth = 2
\makeatother

\onecolumngrid
\begin{center}
 {\large \textmd{Supplemental Materials:} \\[0.3em]
 {\bfseries Multipole Superconductivity in Nonsymmorphic Sr$_2$IrO$_4$}}
\end{center}

\setcounter{page}{1}
\section{Gap classification based on space group symmetry}
\label{sec:gap_classification}
We focus on the magnetic space group of Sr$_2$IrO$_4$ in the $-++-$ state, $M_{-++-}$, which is given as a coset decomposition,
\begin{align}
 M_{-++-} &= G_{-++-} + \{\theta | \bm{\tau}\} G_{-++-}, \label{eq:MSG_-++-} \\
 G_{-++-} &= \{E | \bm{0}\} T + \{I | \bm{0}\} T + \{2_z | \bm{\tau}_x + \bm{\tau}_z\} T + \{\sigma_h | \bm{\tau}_x + \bm{\tau}_z\} T \label{eq:SG_-++-_1} \\
 & \qquad + \{2_x | \bm{\tau}_z\} T + \{2_y | \bm{\tau}_x\} T + \{\sigma_x | \bm{\tau}_z\} T + \{\sigma_y | \bm{\tau}_x\} T, \label{eq:SG_-++-_2}
\end{align}
where the translation group $T$ defines a Bravais Lattice, and $\bm{\tau}_x = \frac{a}{2} \bm{e}_a$, $\bm{\tau}_y = \frac{a}{2} \bm{e}_b$, $\bm{\tau}_z = \frac{c}{2} \bm{e}_c$, $\bm{\tau} = \bm{\tau}_x + \bm{\tau}_y + \bm{\tau}_z$ are non-primitive translation vectors.
The notation $\{p | \bm{a}\}$ is a conventional Seitz space group symbol with a point-group operation $p$ and a translation $\bm{a}$.
$M_{-++-}$ is a nonsymmorphic space group since it contains non-primitive translations.
In Eqs.~\eqref{eq:MSG_-++-}-\eqref{eq:SG_-++-_2} the crystal space group I4$_1$/acd is assumed.
If the crystal space group of Sr$_2$IrO$_4$ is I4$_1$/a, however, symmetry operations in Eq.~\eqref{eq:SG_-++-_2} are not included in $G_{-++-}$.

We define $\gamma^{\bm{k}}_{-++-}(m)$ as a small representation of symmetry operations $m \in {\cal M}^{\bm{k}}_{-++-}$, where ${\cal M}^{\bm{k}}_{-++-} \subset M_{-++-}$ is the ``little group'' leaving $\bm{k}$ invariant modulo a reciprocal lattice vector.
$\gamma^{\bm{k}}_{-++-}$ represents the Bloch state with the crystal momentum $\bm{k}$.
In the superconducting state, the zero-momentum Cooper pairs have to be formed between the degenerate states present at $\bm{k}$ and $- \bm{k}$ within the weak-coupling BCS theory.
Therefore, these two states should be connected by some symmetry operations except for an accidentally degenerate case.
As a result, the representation of Cooper pair wave functions $P^{\bm{k}}_{-++-}$ can be constructed from the representations of the Bloch state $\gamma^{\bm{k}}_{-++-}$.

Here we consider the Cooper pairs on the BPs $k_{z, x, y} = 0$ and the ZFs $k_z = \pm \pi / c$ and $k_{x, y} = \pm \pi / a$.
On each plane, the little group ${\cal M}^{\bm{k}}_{-++-}$ of $M_{-++-}$ is given by the following coset decomposition,
\begin{equation}
 {\cal M}^{\bm{k}}_{-++-} = \begin{cases}
                             \{E | \bm{0}\} T + \{\sigma_h | \bm{\tau}_x + \bm{\tau}_z\} T + \{\theta I | \bm{\tau}\} T + \{\theta 2_z | \bm{\tau}_y\} T & \text{(a) $k_z = 0, \, \pm \pi / c$}, \\
                             \{E | \bm{0}\} T + \{\sigma_x | \bm{\tau}_z\} T + \{\theta I | \bm{\tau}\} T + \{\theta 2_x | \bm{\tau}_x + \bm{\tau}_y\} T & \text{(b) $k_x = 0, \, \pm \pi / a$}, \\
                             \{E | \bm{0}\} T + \{\sigma_y | \bm{\tau}_x\} T + \{\theta I | \bm{\tau}\} T + \{\theta 2_y | \bm{\tau}_y + \bm{\tau}_z\} T & \text{(c) $k_y = 0, \, \pm \pi / a$}.
                            \end{cases} 
\end{equation}
Instead of obtaining the small representations $\gamma^{\bm{k}}_{-++-}$, we calculate the projective IRs $\bar{\gamma}^{\bm{k}}_{-++-}$ of the little co-groups $\bar{\cal M}^{\bm{k}}_{-++-} = {\cal M}^{\bm{k}}_{-++-} / T$ with the appropriate factor systems~\cite{Bradley_SM}.
In Table~\ref{tab:character_gamma_bar}, we summarize the characters of $\bar{\gamma}^{\bm{k}}_{-++-}$ for the unitary operations in $\bar{\cal M}^{\bm{k}}_{-++-}$. Note that the corresponding small representations are given by $\gamma^{\bm{k}}_{-++-}(g) = \bar{\gamma}^{\bm{k}}_{-++-}(r) F^{\bm{k}}(t)$ where $g = rt$ for $g \in {\cal M}^{\bm{k}}_{-++-}$ and $t \in T$. $F^{\bm{k}}$ is the IR of $T$ defined by $F^{\bm{k}}(t) = e^{- i \bm{k} \cdot \bm{t}}$ for $t = \{E | \bm{t}\}$.

\begin{table}[htbp]
 \centering
 \caption{The character of $\bar{\gamma}^{\bm{k}}_{-++-}$. Signs of characters on $k_{x, y} = \pm \pi / a$ correspond to the two non-equivalent IRs.}
 \label{tab:character_gamma_bar}
 \begin{tabular}{c|cc}
  \multicolumn{3}{c}{(a) $k_z = 0, \, \pm \pi / c$} \\ \hline\hline
  $\bar{\cal M}^{\bm{k}}_{-++-}$ & $\{E | \bm{0}\}$ & $\{\sigma_h | \bm{\tau}_x + \bm{\tau}_z\}$ \\ \hline
  BP, ZF & $2$ & $0$ \\ \hline\hline
 \end{tabular}
 \quad
 \begin{tabular}{c|cc}
  \multicolumn{3}{c}{(b) $k_x = 0, \, \pm \pi / a$} \\ \hline\hline
  $\bar{\cal M}^{\bm{k}}_{-++-}$ & $\{E | \bm{0}\}$ & $\{\sigma_x | \bm{\tau}_z\}$ \\ \hline
  BP & $2$ & $0$ \\
  ZF & $2$ & $\pm 2 i e^{- i k_z c / 2}$ \\ \hline\hline
 \end{tabular}
 \quad
 \begin{tabular}{c|cc}
  \multicolumn{3}{c}{(c) $k_y = 0, \, \pm \pi / a$} \\ \hline\hline
  $\bar{\cal M}^{\bm{k}}_{-++-}$ & $\{E | \bm{0}\}$ & $\{\sigma_y | \bm{\tau}_x\}$ \\ \hline
  BP & $2$ & $0$ \\
  ZF & $2$ & $\pm 2 i e^{- i k_x a / 2}$ \\ \hline\hline
 \end{tabular}
\end{table}

Next, we calculate the representation of the Cooper pair wave functions $P^{\bm{k}}_{-++-}$.
Let us consider the space group operation $d = \{p_d | \bm{a}_d\}$ where $p_d$ satisfies $p_d \bm{k} = - \bm{k}$ modulo a reciprocal lattice vector.
The operation $d$ connects two states of the paired electrons.
In the present case, the candidates for the operator $d$ are given by
\begin{equation}
 d = \begin{cases}
      \{I | \bm{0}\}, \, \{2_z | \bm{\tau}_x + \bm{\tau}_z\}, \, \{\theta | \bm{\tau}\}, \, \{\theta \sigma_h | \bm{\tau}_y\} & \text{(a)}, \\
      \{I | \bm{0}\}, \, \{2_x | \bm{\tau}_z\}, \, \{\theta | \bm{\tau}\}, \, \{\theta \sigma_x | \bm{\tau}_x + \bm{\tau}_y\} & \text{(b)}, \\
      \{I | \bm{0}\}, \, \{2_y | \bm{\tau}_x\}, \, \{\theta | \bm{\tau}\}, \, \{\theta \sigma_y | \bm{\tau}_y + \bm{\tau}_z\} & \text{(c)}.
     \end{cases}
\end{equation}
$\widetilde{\cal M}^{\bm{k}}_{-++-} = {\cal M}^{\bm{k}}_{-++-} + d {\cal M}^{\bm{k}}_{-++-}$ is independent of the choice of $d$.
Taking into account the antisymmetry of the Cooper pairs and the degeneracy of the two states, we can regard $P^{\bm{k}}_{-++-}$ as an antisymmetrized Kronecker square~\cite{Bradley_SM, Bradley1970_SM}, with zero total momentum, of the induced representation $\gamma^{\bm{k}}_{-++-} \uparrow \widetilde{\cal M}^{\bm{k}}_{-++-}$.
This is obtained in the systematic way by using the double coset decomposition and the corresponding Mackey-Bradley theorem~\cite{Bradley_SM, Bradley1970_SM, Mackey1953_SM},
\begin{align}
 \chi[P^{\bm{k}}_{-++-}(m)] &= \chi[\gamma^{\bm{k}}_{-++-}(m)] \chi[\gamma^{\bm{k}}_{-++-}(d^{- 1} m d)], \\
 \chi[P^{\bm{k}}_{-++-}(d m)] &= - \chi[\gamma^{\bm{k}}_{-++-}(d m d m)],
\end{align}
where $\chi$ are the characters of the representation.
The obtained results are summarized in Table~\ref{tab:character_P_bar}.
Here, $\bar{P}^{\bm{k}}_{-++-}$ is the representation of $\widetilde{\cal M}^{\bm{k}}_{-++-} / T$ to meet $P^{\bm{k}}_{-++-}(g) = \bar{P}^{\bm{k}}_{-++-}(r)$ where $g = rt$ for $g \in \widetilde{\cal M}^{\bm{k}}_{-++-}$, $r \in \widetilde{\cal M}^{\bm{k}}_{-++-} / T$, and $t \in T$.

\begin{table}[htbp]
 \centering
 \caption{The character of $\bar{P}^{\bm{k}}_{-++-}$.}
 \label{tab:character_P_bar}
 \begin{tabular}{c|cccc}
  \multicolumn{5}{c}{(a) $k_z = 0, \, \pm \pi / c$} \\ \hline\hline
  $\widetilde{\cal M}^{\bm{k}}_{-++-} / T$ & $\{E | \bm{0}\}$ & $\{\sigma_h | \bm{\tau}_x + \bm{\tau}_z\}$ & $\{I | \bm{0}\}$ & $\{2_z | \bm{\tau}_x + \bm{\tau}_z\}$ \\ \hline
  BP & $4$ & $0$ & $- 2$ & $2$ \\
  ZF & $4$ & $0$ & $- 2$ & $- 2$ \\ \hline\hline
 \end{tabular} \\[5mm]
 \begin{tabular}{c|cccc}
  \multicolumn{5}{c}{(b) $k_x = 0, \, \pm \pi / a$} \\ \hline\hline
  $\widetilde{\cal M}^{\bm{k}}_{-++-} / T$ & $\{E | \bm{0}\}$ & $\{\sigma_x | \bm{\tau}_z\}$ & $\{I | \bm{0}\}$ & $\{2_x | \bm{\tau}_z\}$ \\ \hline
  BP & $4$ & $0$ & $- 2$ & $2$ \\
  ZF & $4$ & $- 4$ & $- 2$ & $2$ \\ \hline\hline
 \end{tabular}
 \quad
 \begin{tabular}{c|cccc}
  \multicolumn{5}{c}{(c) $k_y = 0, \, \pm \pi / a$} \\ \hline\hline
  $\widetilde{\cal M}^{\bm{k}}_{-++-} / T$ & $\{E | \bm{0}\}$ & $\{\sigma_y | \bm{\tau}_x\}$ & $\{I | \bm{0}\}$ & $\{2_y | \bm{\tau}_x\}$ \\ \hline
  BP & $4$ & $0$ & $- 2$ & $2$ \\
  ZF & $4$ & $- 4$ & $- 2$ & $2$ \\ \hline\hline
 \end{tabular}
\end{table}

Finally, we reduce the representation $\bar{P}^{\bm{k}}_{-++-}$ into IRs.
In any planes, we have four IRs, $A_g$, $B_g$, $A_u$, and $B_u$ since the coset group $\widetilde{\cal M}^{\bm{k}}_{-++-} / T$ is isomorphic to the gray point group $C_{2h}$.
Then, $\bar{P}^{\bm{k}}_{-++-}$ can be induced to the gray point group $M_{-++-} / T \simeq D_{4h}$ with the help of the ``Frobenius reciprocity theorem''~\cite{Bradley_SM}.
The induced representation $\bar{P}^{\bm{k}}_{-++-} \uparrow M_{-++-} / T$ are summarized in the followings:
\begin{enumerate}
 \item $k_z = 0, \, \pm \pi / c$
       \begin{equation}
        \bar{P}^{\bm{k}}_{-++-} \uparrow M_{-++-} / T =
        \begin{cases}
         \begin{aligned}
          A_{1g} &+ A_{2g} + B_{1g} + B_{2g} + 2 A_{1u} \\
          &+ 2 A_{2u} + 2 B_{1u} + 2 B_{2u} + 2 E_u
         \end{aligned}
         & \text{BP} \\
         2 E_g + A_{1u} + A_{2u} + B_{1u} + B_{2u} + 4 E_u & \text{ZF}
        \end{cases}
       \end{equation}
 \item[(b, c)] $k_{x, y} = 0, \, \pm \pi / a$
       \begin{equation}
        \bar{P}^{\bm{k}}_{-++-} \uparrow M_{-++-} / T =
        \begin{cases}
         \begin{aligned}
          A_{1g} &+ B_{1g} + E_g + 2 A_{1u} \\
          &+ A_{2u} + 2 B_{1u} + B_{2u} + 3 E_u
         \end{aligned}
         & \text{BP} \\
         A_{2g} + B_{2g} + E_g + 3 A_{1u} + 3 B_{1u} + 3 E_u & \text{ZF}
        \end{cases}
       \end{equation}
\end{enumerate}
These results are shown in Eqs.~\eqref{eq:gap_classification_horizontal} and \eqref{eq:gap_classification_vertical}

Here we comment on the case of I4$_1$/a group.
In this case, gap classification shown above is not applicable to the vertical planes $k_{x, y} = 0$ and $k_{x, y} = \pm \pi / a$.
The results of the horizontal planes $k_z = 0, \, \pm \pi / c$ hold in both space groups.

\section{Model}
\label{sec:model}
In this section, we introduce a three-dimensional single-orbital tight-binding model describing superconductivity coexisting with magnetic order in Sr$_2$IrO$_4$,
\begin{equation}
 {\cal H} = \frac{1}{2} \sum_{\bm{k}} \hat{C}_{\bm{k}}^\dagger \hat{H}_{\text{BdG}}(\bm{k}) \hat{C}_{\bm{k}},
\end{equation}
where
\begin{equation}
 \begin{split}
  \hat{C}_{\bm{k}}^\dagger = ( & a_{- / \bm{k}_+ \uparrow}^\dagger, a_{- / \bm{k}_+ \downarrow}^\dagger, \dots, d_{- / \bm{k}_+ \uparrow}^\dagger, d_{- / \bm{k}_+ \downarrow}^\dagger, a_{+ / \bm{k}_+ \uparrow}^\dagger, a_{+ / \bm{k}_+ \downarrow}^\dagger, \dots, d_{+ / \bm{k}_+ \uparrow}^\dagger, d_{+ / \bm{k}_+ \downarrow}^\dagger, \\
  & a_{- / \bm{k}_- \uparrow}, a_{- / \bm{k}_- \downarrow}, \dots, d_{- / \bm{k}_- \uparrow}, d_{- / \bm{k}_- \downarrow}, a_{+ / \bm{k}_- \uparrow}, a_{+ / \bm{k}_- \downarrow}, \dots, d_{+ / \bm{k}_- \uparrow}, d_{+ / \bm{k}_- \downarrow}),
 \end{split} 
\end{equation}
with $\bm{k}_+ \equiv \bm{k} + \frac{\bm{q}}{2}$, $\bm{k}_- \equiv - \bm{k} + \frac{\bm{q}}{2}$ are 32-dimensional vector of creation-annihilation operators.
The center-of-mass momentum $\bm{q}$ of Cooper pairs is assumed to be zero in most cases except for the studies of FFLO state.
We define $a_{\pm / \bm{k} s}, \dots d_{\pm / \bm{k} s}$ as the annihilation operators of electrons with spin $s = \uparrow, \downarrow$ on the sublattices $a_{\pm}, \dots, d_{\pm}$, respectively (see Fig.~\ref{fig:Sr2IrO4} for the sublattices)
The $32 \times 32$ BdG Hamiltonian is described with use of the normal state Hamiltonian $\hat{H}_{n}(\bm{k})$ and the order parameter part $\hat{\Delta}(\bm{k})$,
\begin{equation}
 \hat{H}_{\text{BdG}}(\bm{k}) =
 \begin{pmatrix}
  \hat{H}_{n}(\bm{k}_+) & \hat{\Delta}(\bm{k}) \\
  \hat{\Delta}(\bm{k})^\dagger & - \hat{H}_{n}^{\text{T}}(\bm{k}_-)
 \end{pmatrix},
 \label{eq:BdG_Hamiltonian}
\end{equation}
where
\begin{equation}
 \hat{H}_{n}(\bm{k}) = \hat{H}_{\text{kin}}(\bm{k}) + \hat{H}_{\text{ASOC}}(\bm{k}) + \hat{H}_{\text{MO}}.
  \label{eq:Hamiltonian_normal}
\end{equation}
The kinetic term $\hat{H}_{\text{kin}}(\bm{k})$ is given by the following equation:
\begin{equation}
 \hat{H}_{\text{kin}}(\bm{k}) = \left[
                                 \begin{array}{c|c}
                                  \begin{tabular}{l}
                                   $\hat{H}_{\text{intra-layer}}(\bm{k})$ \\
                                   $+ \hat{H}_{\text{inter-layer1}}(\bm{k})$ \\
                                  \end{tabular} & 
                                  \begin{tabular}{l}
                                   $\hat{H}_{\text{inter-layer2}}(\bm{k})$ \\
                                  \end{tabular} \\ \hline
                                  \begin{tabular}{l}
                                   $\hat{H}_{\text{inter-layer2}}(\bm{k})^\dagger$ \\
                                  \end{tabular} &
                                  \begin{tabular}{l}
                                   $\hat{H}_{\text{intra-layer}}(\bm{k})$ \\
                                   $+ \hat{H}_{\text{inter-layer1}}(\bm{k})$ \\
                                  \end{tabular}
                                 \end{array} \right],
\end{equation}
where
\begin{align}
 \hat{H}_{\text{intra-layer}}(\bm{k}) &= \hat{\sigma}_0^{\text{(layer)}} \otimes [ (\varepsilon_2(\bm{k}) - \mu) \hat{\sigma}_0^{\text{(sl)}} \otimes \hat{\sigma}_0^{\text{(spin)}} + \varepsilon_1(\bm{k}) \hat{\sigma}_x^{\text{(sl)}} \otimes \hat{\sigma}_0^{\text{(spin)}} ], \\
 \hat{H}_{\text{inter-layer1}}(\bm{k}) &= \hat{\sigma}_x^{\text{(layer)}} \otimes [ \Re(\varepsilon_3^x(\bm{k})) \hat{\sigma}_0^{\text{(sl)}} \otimes \hat{\sigma}_0^{\text{(spin)}} + \Re(\varepsilon_3^y(\bm{k})) \hat{\sigma}_x^{\text{(sl)}} \otimes \hat{\sigma}_0^{\text{(spin)}} ] \notag \\
 & \qquad - \hat{\sigma}_y^{\text{(layer)}} \otimes [ \Im(\varepsilon_3^x(\bm{k})) \hat{\sigma}_0^{\text{(sl)}} \otimes \hat{\sigma}_0^{\text{(spin)}} + \Im(\varepsilon_3^y(\bm{k})) \hat{\sigma}_x^{\text{(sl)}} \otimes \hat{\sigma}_0^{\text{(spin)}} ], \\
 \hat{H}_{\text{inter-layer2}}(\bm{k}) &= \hat{\sigma}_x^{\text{(layer)}} \otimes [ \Re(\varepsilon_3^y(\bm{k})) \hat{\sigma}_0^{\text{(sl)}} \otimes \hat{\sigma}_0^{\text{(spin)}} + \Re(\varepsilon_3^x(\bm{k})) \hat{\sigma}_x^{\text{(sl)}} \otimes \hat{\sigma}_0^{\text{(spin)}} ] \notag \\
 & \qquad + \hat{\sigma}_y^{\text{(layer)}} \otimes [ \Im(\varepsilon_3^y(\bm{k})) \hat{\sigma}_0^{\text{(sl)}} \otimes \hat{\sigma}_0^{\text{(spin)}} + \Im(\varepsilon_3^x(\bm{k})) \hat{\sigma}_x^{\text{(sl)}} \otimes \hat{\sigma}_0^{\text{(spin)}} ],
\end{align}
with the chemical potential $\mu$.
$\hat{\sigma}_i^{\text{(spin)}}$, $\hat{\sigma}_i^{\text{(sl)}}$, and $\hat{\sigma}_i^{\text{(layer)}}$ are the Pauli matrices representing the spin, sublattice, and layer degrees of freedom, respectively.
The single electron kinetic energy terms $\varepsilon_1(\bm{k})$, $\varepsilon_2(\bm{k})$, and $\varepsilon_3^{x, y}(\bm{k})$ are described by taking into account the nearest-, next-nearest-, and third-nearest-neighbor hoppings,
\begin{align}
 \varepsilon_1(\bm{k}) &= - 4 t_1 \cos\frac{k_x a}{2} \cos\frac{k_y a}{2}, \\
 \varepsilon_2(\bm{k}) &= - 2 t_2 (\cos(k_x a) + \cos(k_y a)), \\
 \varepsilon_3^x(\bm{k}) &= - t_3 \cos\frac{k_x a}{2} e^{- i k_z c / 4}, \\
 \varepsilon_3^y(\bm{k}) &= - t_3 \cos\frac{k_y a}{2} e^{- i k_z c / 4}.
\end{align}

For our results in the $-+-+$ state, the violation of local inversion symmetry which induces the staggered ASOC, $\hat{H}_{\text{ASOC}}(\bm{k})$, plays an essential role.
This term is given by the following matrix:
\begin{equation}
 \hat{H}_{\text{ASOC}}(\bm{k}) = \left[
                                  \begin{array}{c|c}
                                   \begin{tabular}{l}
                                    $\hat{H}_{\text{ASOC-intra1}}(\bm{k})$ \\
                                    $+ \hat{H}_{\text{ASOC-intra2}}(\bm{k})$ \\
                                    $+ \hat{H}_{\text{ASOC-inter}}^y(\bm{k})$ \\
                                   \end{tabular} & 
                                   \begin{tabular}{l}
                                    $\hat{H}_{\text{ASOC-inter}}^x(\bm{k})$ \\
                                   \end{tabular} \\ \hline
                                   \begin{tabular}{l}
                                    $\hat{H}_{\text{ASOC-inter}}^x(\bm{k})^\dagger$ \\
                                   \end{tabular} &
                                   \begin{tabular}{l}
                                    $\hat{H}_{\text{ASOC-intra1}}(\bm{k})$ \\
                                    $+ \hat{H}_{\text{ASOC-intra2}}(\bm{k})$ \\
                                    $+ \hat{H}_{\text{ASOC-inter}}^y(\bm{k})$ \\
                                   \end{tabular}
                                  \end{array} \right].
\end{equation}
We take into account two intra-layer terms $\hat{H}_{\text{ASOC-intra1}}(\bm{k}), \hat{H}_{\text{ASOC-intra2}}(\bm{k})$ and two inter-layer terms $\hat{H}_{\text{ASOC-inter}}^{x, y}(\bm{k})$:
\begin{align}
 \hat{H}_{\text{ASOC-intra1}}(\bm{k}) &= i \alpha_1 \cos\frac{k_x a}{2} \cos\frac{k_y a}{2} \hat{\sigma}_0^{\text{(layer)}} \otimes i \hat{\sigma}_y^{\text{(sl)}} \otimes \hat{\sigma}_z^{\text{(spin)}}, \\
 \hat{H}_{\text{ASOC-intra2}}(\bm{k}) &= \alpha_2 \hat{\sigma}_z^{\text{(layer)}} \otimes \hat{\sigma}_z^{\text{(sl)}} \otimes ( \sin(k_x a) \cos(k_y a) \hat{\sigma}_x^{\text{(spin)}} - \sin(k_y a) \cos(k_x a) \hat{\sigma}_y^{\text{(spin)}} ), \\
 \hat{H}_{\text{ASOC-inter}}^x(\bm{k}) &= - \alpha_3 \left[ i \hat{\sigma}_y^{\text{(layer)}} \otimes i \hat{\sigma}_y^{\text{(sl)}} \otimes \left( \cos\frac{k_z c}{4} \sin\frac{k_x a}{2} \hat{\sigma}_x^{\text{(spin)}} - 2 \sin\frac{k_z c}{4} \cos\frac{k_x a}{2} \hat{\sigma}_z^{\text{(spin)}} \right) \right. \notag \\
 & \qquad \left. + i \cdot \hat{\sigma}_x^{\text{(layer)}} \otimes i \hat{\sigma}_y^{\text{(sl)}} \otimes \left( \sin\frac{k_z c}{4} \sin\frac{k_x a}{2} \hat{\sigma}_x^{\text{(spin)}} + 2 \cos\frac{k_z c}{4} \cos\frac{k_x a}{2} \hat{\sigma}_z^{\text{(spin)}} \right) \right], \\
 \hat{H}_{\text{ASOC-inter}}^y(\bm{k}) &= \alpha_3 \left[ i \hat{\sigma}_y^{\text{(layer)}} \otimes i \hat{\sigma}_y^{\text{(sl)}} \otimes \left( \cos\frac{k_z c}{4} \sin\frac{k_y a}{2} \hat{\sigma}_y^{\text{(spin)}} - 2 \sin\frac{k_z c}{4} \cos\frac{k_y a}{2} \hat{\sigma}_z^{\text{(spin)}} \right) \right. \notag \\
 & \qquad \left. - i \cdot \hat{\sigma}_x^{\text{(layer)}} \otimes i \hat{\sigma}_y^{\text{(sl)}} \otimes \left( \sin\frac{k_z c}{4} \sin\frac{k_y a}{2} \hat{\sigma}_y^{\text{(spin)}} + 2 \cos\frac{k_z c}{4} \cos\frac{k_y a}{2} \hat{\sigma}_z^{\text{(spin)}} \right) \right],
\end{align}
which are allowed by the crystal symmetry of Sr$_2$IrO$_4$.

The last term in Eq.~\eqref{eq:Hamiltonian_normal}, $\hat{H}_{\text{MO}}$, expresses the molecular field of magnetic order, $-++-$ and $-+-+$.
This term causes various superconducting phenomena, which have been demonstrated in this paper.
As shown in Fig.~\ref{fig:Sr2IrO4}, each site has the in-plane magnetic moment.
Thus, the molecular field is given by
\begin{equation}
 \hat{H}_{\text{MO}} =
  \begin{bmatrix}
   - \bm{h}(\theta_{a -}) \cdot \bm{\hat{\sigma}} & & \\
   & \ddots & \\
   & & - \bm{h}(\theta_{d +}) \cdot \bm{\hat{\sigma}}
  \end{bmatrix},
\end{equation}
where
\begin{equation}
 (\theta_{a -}, \dots, \theta_{d -}, \theta_{a +}, \dots, \theta_{d +})
  = \begin{cases}
     (348^\circ, 192^\circ, 168^\circ, 12^\circ, 168^\circ, 12^\circ, 348^\circ, 192^\circ) & (\text{$-++-$ state}) \\
     (348^\circ, 192^\circ, 168^\circ, 12^\circ, 348^\circ, 192^\circ, 168^\circ, 12^\circ) & (\text{$-+-+$ state}),
    \end{cases}
\end{equation}
and $\bm{h}(\theta) = h (\cos\theta, \sin\theta, 0)$~\cite{Matteo2016_SM}.

Next we describe the order parameter $\hat{\Delta}(\bm{k})$.
When the on-site $s$-wave superconductivity is assumed, it takes the form
\begin{equation}
 \hat{\Delta}^{(s)}(\bm{k}) = \Delta_0 \hat{1}_2 \otimes \hat{\sigma}_0^{\text{(layer)}} \otimes \hat{\sigma}_0^{\text{(sl)}} \otimes i \hat{\sigma}_y^{\text{(spin)}}.
  \label{eq:s-wave_order_parameter}
\end{equation}
For the $d_{xy}$-wave superconductivity originating from the interaction between the nearest-neighbor sites, we obtain
\begin{equation}
 \hat{\Delta}^{(d)}(\bm{k}) = \Delta_0 \sin\frac{k_x a}{2} \sin\frac{k_y a}{2} \hat{1}_2 \otimes \hat{\sigma}_0^{\text{(layer)}} \otimes \hat{\sigma}_x^{\text{(sl)}} \otimes i \hat{\sigma}_y^{\text{(spin)}}.
  \label{eq:d-wave_order_parameter}
\end{equation}

Finally, we show the parameters which are used in this paper.
We adopt the hopping parameters of the effective $J_{\text{eff}} = 1 / 2$ model~\cite{HiroshiWatanabe2013_SM} derived from the three-orbital Hubbard model, where the hopping parameters are $t_1 = 1$, $t_2 = 0.26$, and $t_3 = 0.1$.
We here assume moderate ASOCs $\alpha_1 = 0.3$ and $\alpha_2 = \alpha_3 = 0.1$ so that the effects of ASOCs are visible in the numerical results.
Since the superconductivity has been predicted at the electron density around $n \sim 1.2$~\cite{HiroshiWatanabe2013_SM}, we determine the chemical potential $\mu = 1.05$ so as to be consistent with the electron density.
Then, four spinful energy bands cross the Fermi level.
The magnitude of gap function is chosen to be $\Delta_0 = 0.02$.
The conclusions of this paper are not altered by the choice of parameters, because they are evidenced by the group theoretical analysis.

\section{Accidental gap of $A_{1g}$ state at $k_y = \pm \pi / a$ in $-++-$ state}
\label{sec:accidental_gap}
\begin{figure}[htbp]
 \centering
 \includegraphics[width=12cm, clip]{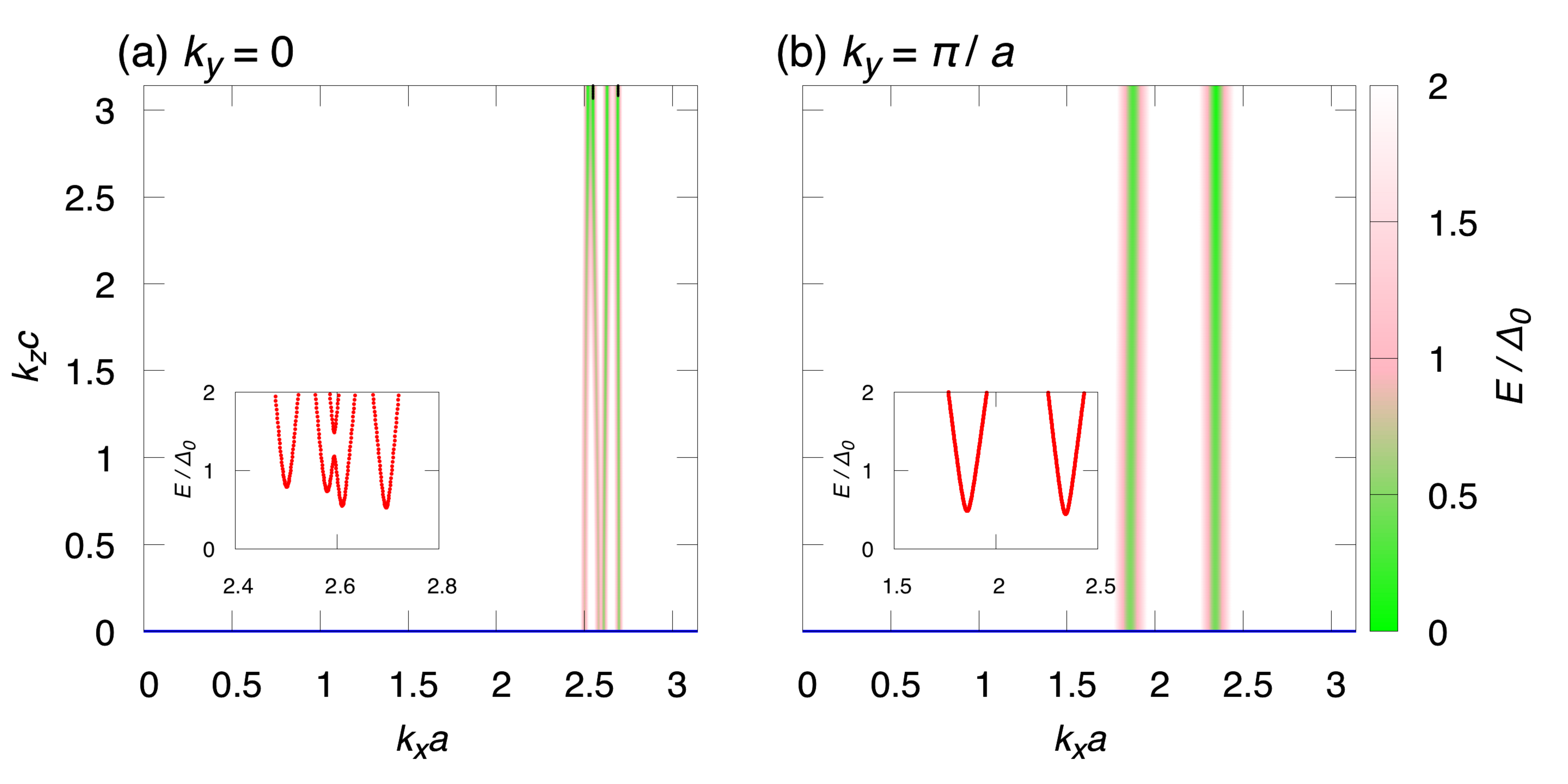}
 \caption{(Color online) The contour plot of quasiparticle energy dispersion $E$ in the $s$-wave superconducting state normalized by the order parameter $\Delta_0$ on (a) $k_y = 0$ and (b) $k_y = \pm \pi / a$. The insets in (a) and (b) show the dispersion $E / \Delta_0$ along the respective blue line. (a) On the BP $k_y = 0$, quasiparticle in almost whole region except for on the BZ boundary $k_z = \pi / c$ are gapped. This is consistent with the gap classification. (b) On the ZF $k_y = \pm \pi / a$, line nodes vanish in disagreement with the gap classification.}
 \label{fig:mppm_s-wave_vertical_ky}
\end{figure}

Gap classification using the space group symmetry reveals that the $A_{1g}$ gap functions in the $-++-$ state possess vertical line nodes on the ZF $k_y = \pm \pi / a$, although the $A_{1g}$ representation is allowed on the BP $k_y = 0$.
In our numerical calculation, however, a small gap appears in the excitation spectrum on the ZF although the magnitude of the gap is smaller than that on BP (see Fig.~\ref{fig:mppm_s-wave_vertical_ky}).
That is because single-particle states are accidentally fourfold degenerate all over the ZF $k_y = \pm \pi / a$ in our model.
This fourfold degeneracy is not protected by symmetry except for on some high-symmetry lines (Sec.~\ref{sec:fourfold_degeneracy}).
The group theoretical analysis of gap classification can be applied only to the intra-band gap, which are diagonal components of the band-based order parameter matrix~\cite{Yanase2016_SM, Nomoto2017_SM, Micklitz2017_SM}.
In ordinary cases, intra-band gap is equivalent to the excitation gap since inter-band gap (offdiagonal components of the band-based order parameter matrix) hardly affects the energy spectrum near $E = 0$.
In the presence of (nearly) fourfold degeneracy, however, inter-band gap may induce excitation gap~\cite{Yanase2016_SM}.
Then, the gap nodes expected from the gap classification can be lost.
Indeed, such a gap opening changes the nodal line to nodal loops in UPt$_3$~\cite{Yanase2016_SM}.
In many cases including UPt$_3$, however, the inter-band gap appears only on the high-symmetry lines, and the dimension of nodes is not altered.
Our tight-binding model accidentally has fourfold degeneracy on the plane, and therefore, we obtain the excitation gap on the ZF $k_y = \pm \pi / a$.
We believe that the gap at $k_y = \pm \pi / a$ is lifted by taking into account all the spin-orbit couplings allowed by the symmetry.

\section{Symmetry-protected Dirac line nodes on BZ boundary in $-++-$ state}
\label{sec:fourfold_degeneracy}
\begin{figure}[htbp]
 \centering
 \includegraphics[width=6cm, clip]{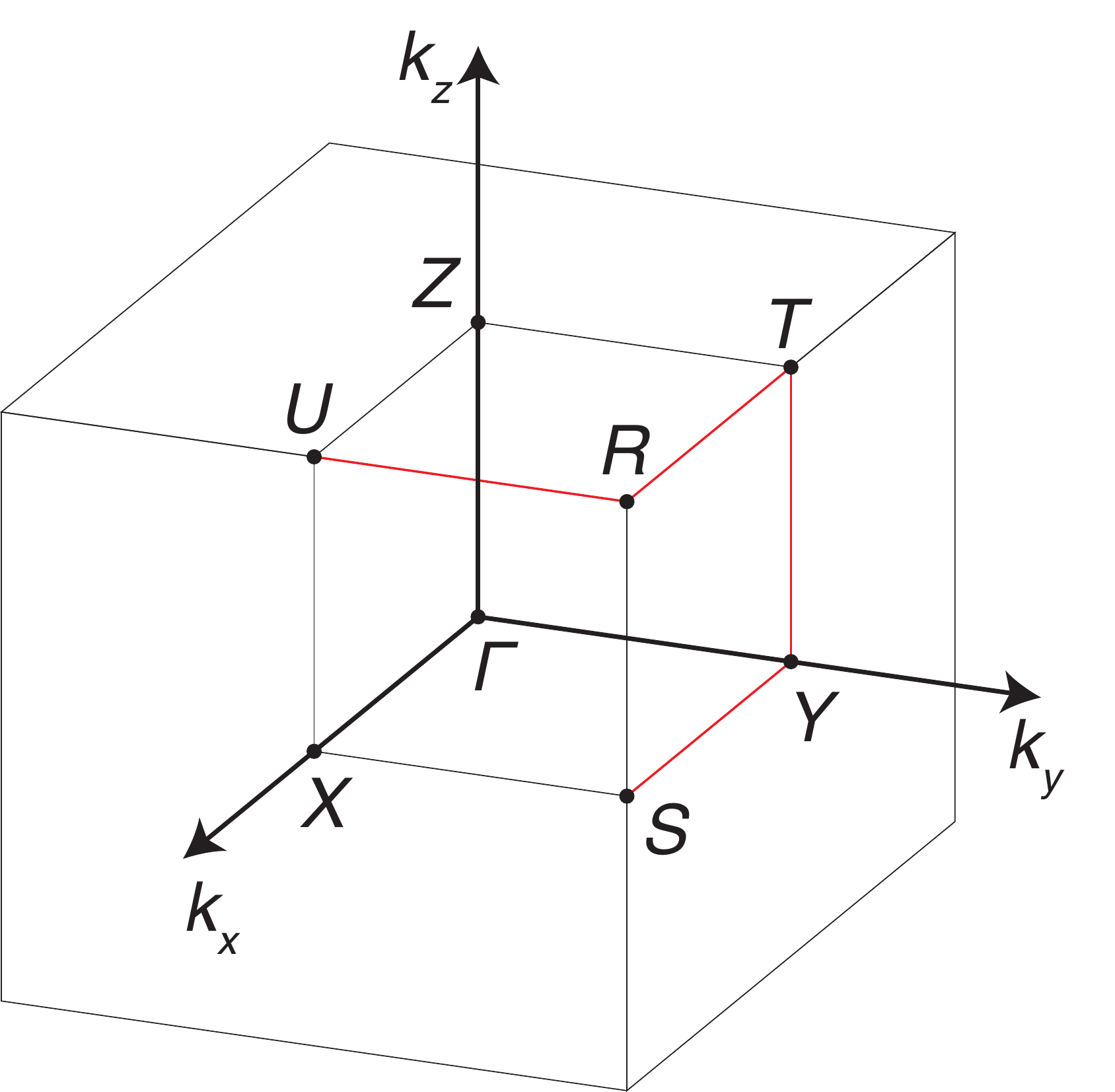}
 \caption{(Color online) The first BZ for primitive orthorhombic lattice. Single particle states are fourfold degenerate on the red lines.}
 \label{fig:primitive_orthorhombic}
\end{figure}

We show the symmetry protection of the fourfold degeneracy on the BZ boundary in $-++-$ state.
The fourfold degeneracy appears at $U$-$R$, $R$-$T$, $T$-$Y$, and $Y$-$S$ lines in the first BZ (Fig.~\ref{fig:primitive_orthorhombic}).
Using the little group on each line, we prove the presence of the degeneracy by symmetry.

On the $U$-$R$ line ($k_y = \pm \pi / a$ and $k_z = \pm \pi / c$), the little group is given by
\begin{align}
 & \{E | \bm{0}\} T + \{\sigma_h | \bm{\tau}_x + \bm{\tau}_z\} T + \{\sigma_x | \bm{\tau}_z\} T + \{2_y | \bm{\tau}_x\} T \notag \\
 & \quad + \{\theta I | \bm{\tau}\} T + \{\theta 2_z | \bm{\tau}_y\} T + \{\theta 2_x | \bm{\tau}_x + \bm{\tau}_y\} T + \{\theta \sigma_y | \bm{\tau}_y + \bm{\tau}_z\} T.
\end{align}
The fourfold degeneracy is proven from algebra, $(\{2_y | \bm{\tau}_x\})^2 = - 1$, $\bigl\{ \{2_y | \bm{\tau}_x\}, \{\sigma_x | \bm{\tau}_z\} \bigr\} = 0$, and $\bigl\{ \{2_y | \bm{\tau}_x\}, \{\theta I | \bm{\tau}\} \bigr\} = 0$~\cite{Yanase2016_SM, LiangQi2016_SM, Sumita2016_SM}.
Because of the rotation symmetry $\{2_y | \bm{\tau}_x\}$, the normal part Hamiltonian on the $U$-$R$ line is block diagonalized and decomposed into the $\pm i$ subsectors.
The $\{\theta I | \bm{\tau}\}$ symmetry is preserved in each subsector as ensured by the anticommutation relation between $\{2_y | \bm{\tau}_x\}$ and $\{\theta I | \bm{\tau}\}$.
Thus, Kramers pairs are formed in each subsector.
The anticommutation relation between $\{2_y | \bm{\tau}_x\}$ and $\{\sigma_x | \bm{\tau}_z\}$ ensures that a Kramers pair in the $i$ subsector is degenerate with another Kramers pair in the $- i$ subsector.
Thus, the fourfold degeneracy is protected by symmetry.

On the other lines, the fourfold degeneracy is proved in a similar way.
On the $R$-$T$ and $Y$-$S$ lines, we use the relations, $(\{2_x | \bm{\tau}_z\})^2 = - 1$, $\bigl\{ \{2_x | \bm{\tau}_z\}, \{\sigma_y | \bm{\tau}_x\} \bigr\} = 0$, and $\bigl\{ \{2_x | \bm{\tau}_z\}, \{\theta I | \bm{\tau}\} \bigr\} = 0$.
Finally on the $T$-$Y$ line, the fourfold degeneracy is proved by the relations, $(\{\sigma_y | \bm{\tau}_x\})^2 = - 1$, $\bigl\{ \{\sigma_y | \bm{\tau}_x\}, \{2_z | \bm{\tau}_x + \bm{\tau}_z\} \bigr\} = 0$, and $\bigl\{ \{\sigma_y | \bm{\tau}_x\}, \{\theta I | \bm{\tau}\} \bigr\} = 0$.

\section{Classification of $-++-$ and $-+-+$ order based on magnetic multipole}
\label{sec:multipole}
We show that the $-++-$ and $-+-+$ order are classified into a magnetic octupole (MO) and magnetic quadrupole (MQ) order, respectively.

\subsection{$-++-$ order}
Although the crystal symmetry of Sr$_2$IrO$_4$ is $D_{4h}$, it reduces to $D_{2h}$ in the $-++-$ ordered state.
In Table~\ref{tab:multipole_mppm}, the even-parity IRs of $D_{4h}$ except $A_{1g}$ ($A_{2g}$, $B_{1g}$, $B_{2g}$, and $E_g$) are subduced to representations of $D_{2h}$.
Since only $B_{1g}$ contains the fully symmetric IR of $D_{2h}$ ($A_g$), the $-+-+$ order belongs to $B_{1g}$ representation of $D_{4h}$.

\begin{table}[htbp]
 \centering
 \caption{Irreducible decomposition of $D_{4h}$ even-parity IRs in $D_{2h}$ point group.}
 \label{tab:multipole_mppm}
 \begin{tabular}{c|cccc} \hline\hline
  (IRs of $D_{4h}$) & $A_{2g}$ & $B_{1g}$ & $B_{2g}$ & $E_g$ \\ \hline
  (IRs of $D_{4h}$)$\downarrow D_{2h}$ & $B_{1g}$ & $A_g$ & $B_{1g}$ & $B_{2g} + B_{3g}$ \\ \hline\hline
 \end{tabular}
\end{table}

The lowest-order time-reversal-odd basis function of $B_{1g}$ is $\alpha x y \hat{\sigma}_z + \beta z (y \hat{\sigma}_x + x \hat{\sigma}_y)$ in the real space.
This basis function represents an even-parity MO ($l = 3$) order~\cite{Schwartz1955_SM},
\begin{gather}
 \hat{M}_{3, -2} - \hat{M}_{3, 2} \propto x y \hat{z} + y z \hat{x} + z x \hat{y}, \\
 \hat{M}_{l, m} = \mu_B \sum_{j = 1}^{n} \left( \frac{2 \bm{l}_j}{l + 1} + 2 \bm{s}_j \right) \cdot \nabla_j \left( r_j^l Z_{l, m}(\hat{\bm{r}}_j)^* \right),
\end{gather}
where $Z_{l, m}(\hat{\bm{r}}) \equiv \sqrt{\frac{4 \pi}{2l + 1}} Y_{l, m}(\hat{\bm{r}})$ is the normalized spherical harmonics.
Thus, the $-++-$ order is classified into a MO order.

\subsection{$-+-+$ order}
In the $-+-+$ ordered state, the crystal symmetry reduces from $D_{4h}$ to $C_{2v}$ .
Here, the odd-parity IRs of $D_{4h}$ ($A_{1u}$, $A_{2u}$, $B_{1u}$, $B_{2u}$, and $E_u$) are subduced to representations of $C_{2v}$ (Table~\ref{tab:multipole_mpmp}).
Since only $E_u$ contains the fully symmetric IR of $C_{2v}$ ($A_1$), the $-+-+$ order belongs to $E_u$ representation of $D_{4h}$.

\begin{table}[htbp]
 \centering
 \caption{Irreducible decomposition of $D_{4h}$ odd-parity IRs in $C_{2v}$ point group.}
 \label{tab:multipole_mpmp}
 \begin{tabular}{c|ccccc} \hline\hline
  (IRs of $D_{4h}$) & $A_{1u}$ & $A_{2u}$ & $B_{1u}$ & $B_{2u}$ & $E_u$ \\ \hline
  (IRs of $D_{4h}$)$\downarrow C_{2v}$ & $A_2$ & $B_2$ & $A_2$ & $B_2$ & $A_1 + B_1$ \\ \hline\hline
 \end{tabular}
\end{table}

This IR $E_u$ permits time-reversal-odd basis functions: $\alpha y \hat{\sigma}_z + \beta z \hat{\sigma}_y$ in the real space, and $k_x$ in the momentum space.
In the real space, the basis function contains an odd-parity MQ ($l = 2$) order~\cite{Schwartz1955_SM},
\begin{equation}
 \hat{M}_{2, 1} + \hat{M}_{2, -1} \propto y \hat{z} + z \hat{y}.
\end{equation}
Therefore, the $-+-+$ order contains the component of a MQ order, though it may include a toroidal dipole order proportional to $y \hat{z} - z \hat{y}$~\cite{Spaldin2008_SM}.
In the momentum space, the linear $k_x$ function makes the band structure asymmetric along the $k_x$ axis, which is demonstrated in Sec.~\ref{sec:mpmp_band}.

\section{Band structure in $-+-+$ state}
\label{sec:mpmp_band}
\begin{figure}[htbp]
 \centering
 \includegraphics[width=8cm, clip]{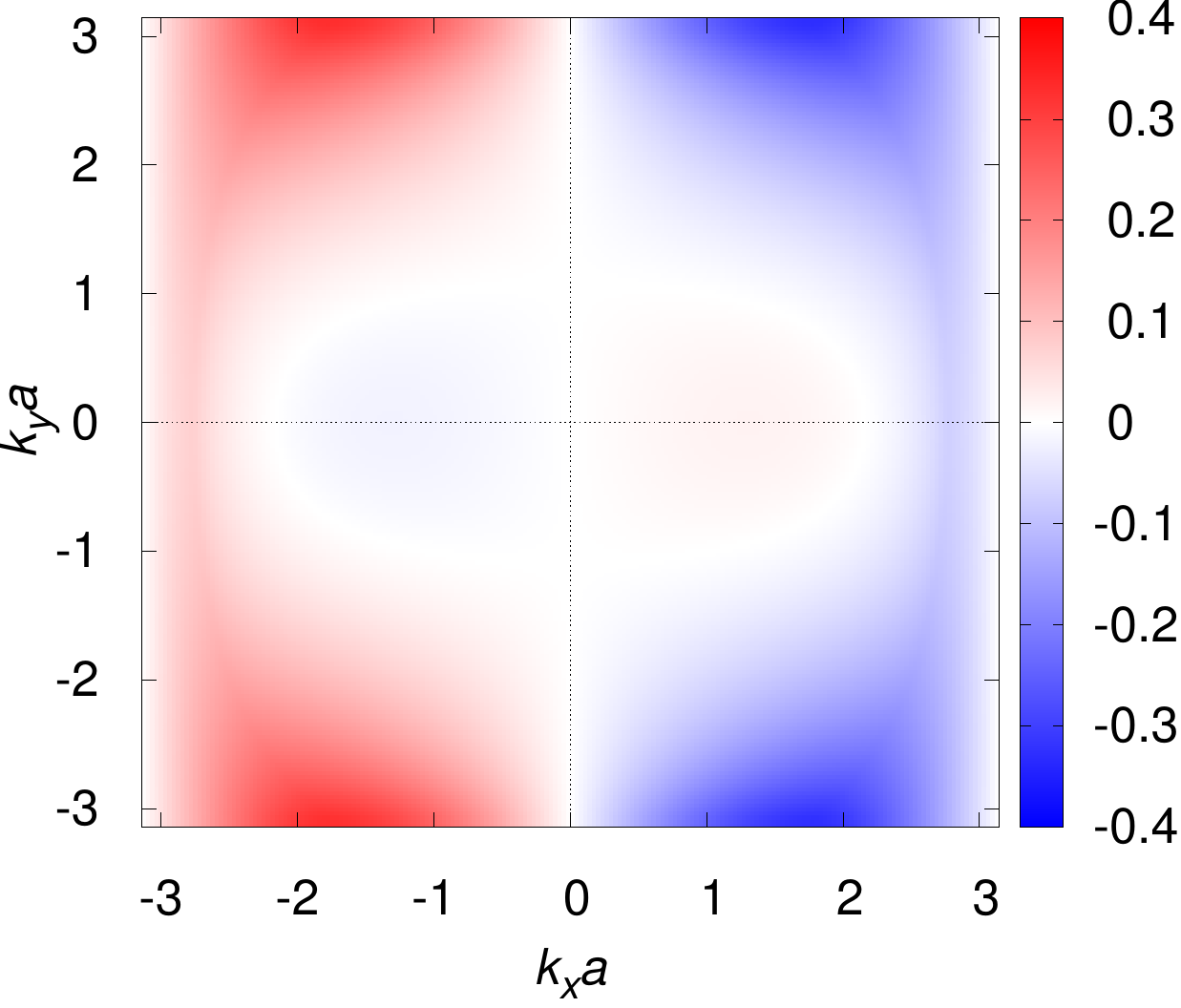}
 \caption{(Color online) The difference of normal energy dispersion, $E_n(k_x, k_y, 0) - E_n(- k_x, k_y, 0)$ which quantifies the band asymmetry. Colored plot shows that the band structure is asymmetric along the $k_x$ axis, while it is symmetric with respect to $k_y$.}
 \label{fig:mpmp_asym}
\end{figure}

As shown in the main text and Sec.~\ref{sec:multipole}, the $-+-+$ magnetic order contains the component of a MQ order which makes the band structure asymmetric along the $k_x$ axis.
We demonstrate the asymmetry using our effective $J_{\text{eff}} = 1 / 2$ model (Sec.~\ref{sec:model}).
Figure~\ref{fig:mpmp_asym} shows the contour plot of $E_n(k_x, k_y, 0) - E_n(- k_x, k_y, 0)$, where $E_n(\bm{k})$ is one of the normal energy dispersions.
The colored region implies the asymmetry along the $k_x$ axis of the band structure.
The asymmetry is particularly pronounced near the BZ boundary, and the Fermi surface of doped Sr$_2$IrO$_4$ is close to the BZ boundary (Figs.~\ref{fig:mppm_s-wave} and \ref{fig:mppm_d-wave}).
Thus, the $-+-+$ magnetic order significantly affects the superconductivity through the band asymmetry.
Moreover, the band structure is obviously symmetric with respect to $k_y$.

These symmetric/asymmetric properties are understood by considering the symmetry operations preserved in the $-+-+$ state.
The system is invariant under the operations which flip the wave number $k_y$ to $- k_y$: the twofold rotation $\{2_x | \bm{\tau}_z\}$, the twofold screw operation $\{2_x | \bm{\tau}_x + \bm{\tau}_y\}$, and the glide operations $\{\sigma_y | \bm{\tau}_x\}$ and $\{\sigma_y | \bm{\tau}_y + \bm{\tau}_z\}$.
The operations which flip the wave number $k_z$ are similarly preserved.
However, the $-+-+$ state is \textit{not} invariant under the operations which flip $k_x$, such as the twofold rotations (screw operations) $2_y, 2_z$, the glide operations $\sigma_x$, and the time-reversal $\theta$.
Namely, all the symmetries protecting the symmetric band structure along the $k_x$ axis are broken.

Then, we also notice a twofold degeneracy in the band structure protected by symmetry.
The $-+-+$ magnetic order spontaneously breaks the inversion symmetry $I$ as well as the time-reversal symmetry $\theta$ in spite of the globally centrosymmetric crystal structure.
However the combined $\theta I$ symmetry is preserved.
This combined operation satisfies $(\theta I)^2 = - 1$ which ensures a twofold degeneracy in the band structure as proved by the Kramers theorem.

Finally we briefly comment on the validity of assuming $s$-wave superconductivity to calculate superconducting susceptibility in the $-+-+$ state.
Regardless of the form of the superconducting order parameter, the fact remains that the band structure asymmetrically deforms in the $-+-+$ state, as shown above.
The asymmetry linear in $k_x$ ensures the FFLO state irrespective of the symmetry of superconducting order parameter.
Therefore, the FFLO state shown in the main text should also be stabilized in the case of unconventional superconductivity.

\section{Calculation of superconducting susceptibility}
\label{sec:susceptibility}
Here we show the definition and calculation of superconducting susceptibility.
We define the susceptibility as,
\begin{equation}
 \chi_{m m'}(\bm{q}, i \Omega_n) = \int_{0}^{\beta} d\tau e^{i \Omega_n \tau} \braket{B_{m}(\bm{q}, \tau) B_{m'}^\dagger(\bm{q}, 0)},
\end{equation}
where $\Omega_n = 2 n \pi T$ is the bosonic Matsubara frequency, and $m = 1, 2, \dots, 8$ represents the sublattice $a_-, \dots, d_-, a_+, \dots, d_+$, respectively.
The creation operator of Cooper pairs has been introduced as
\begin{equation}
 B_{m}^\dagger(\bm{q}) = \frac{1}{\sqrt{2V}} \sum_{\bm{k}, s, s'} (i \hat{\sigma}_y)_{s s'} c_{\bm{k} s m}^\dagger c_{- \bm{k} + \bm{q} s' m}^\dagger,
\end{equation}
where we assume the local $s$-wave superconductivity for simplicity, and $B_{m}(\bm{q}, \tau) = e^{H_{n} \tau} B_{m}(\bm{q}) e^{- H_{n} \tau}$.
$c_{\bm{k} s m}$ is the annihilation operator of electrons with spin $s = \uparrow, \downarrow$ on the sublattice $m$:
\begin{align}
 & (c_{\bm{k} s 1}, c_{\bm{k} s 2}, c_{\bm{k} s 3}, c_{\bm{k} s 4}, c_{\bm{k} s 5}, c_{\bm{k} s 6}, c_{\bm{k} s 7}, c_{\bm{k} s 8}) \notag \\
 & \qquad = (a_{- / \bm{k} s}, b_{- / \bm{k} s}, c_{- / \bm{k} s}, d_{- / \bm{k} s}, a_{+ / \bm{k} s}, b_{+ / \bm{k} s}, c_{+ / \bm{k} s}, d_{+ / \bm{k} s}).
\end{align}

Since it is impossible to exactly calculate the superconducting susceptibility, we apply the $T$-matrix approximation, which is equivalent to the mean-field approximation.
By using the $T$-matrix approximation, the susceptibility matrix $\hat{\chi} = (\chi_{m m'})$ is given by
\begin{equation}
 \hat{\chi}(\bm{q}, i \Omega_n) = \frac{\hat{\chi}^{(0)}(\bm{q}, i \Omega_n)}{\hat{1}_8 - U \hat{\chi}^{(0)}(\bm{q}, i \Omega_n)},
\end{equation}
where $U$ is the $s$-wave on-site attraction.
The irreducible susceptibility $\hat{\chi}^{(0)}$ is given by the following equation:
\begin{equation}
 \chi^{(0)}_{m m'}(\bm{q}, i \Omega_n) = \frac{1}{\beta V} \sum_{\bm{k}} \sum_{s_1, \dots, s_4} \sum_{l} (i \sigma_y)_{s_1 s_2} (i \sigma_y)_{s_3 s_4} G_{m m'}^{s_1 s_3}(\bm{k}, i \omega_l) G_{m m'}^{s_2 s_4}(- \bm{k} + \bm{q}, i \Omega_n - i \omega_l),
\end{equation}
where $G_{m m'}^{s s'}(\bm{k}, i \omega_l)$ is the noninteracting Green's function, and $\omega_l = (2 l + 1) \pi T$ is the fermionic Matsubara frequency.


\begin{thebibliography}{70}
\bibitem [{d-w()}]{d-wave}%
  \BibitemOpen
  \href@noop {} {}\bibinfo {note} {Because of the $\pi / 4$ rotation of a
  principle axis, the $d_{xy}$-wave superconductivity corresponds to the
  $d_{x^2 - y^2}$-wave state in cuprates.}\BibitemShut {Stop}%
\end{thebibliography}

%

\end{document}